\documentclass[doublecol]{epl2}
\usepackage{graphicx,epstopdf}
\usepackage{amssymb}
\title{Chimera states: Effects of different coupling topologies}
\shorttitle{Chimera states: Effects of different coupling topologies}

\author{Bidesh K. Bera,\inst{1} Soumen Majhi,\inst{1} Dibakar Ghosh,\inst{1} and Matja\v{z} Perc\inst{2,3}}
\shortauthor{B. K. Bera, S. Majhi, D. Ghosh, and M. Perc}

\institute{\inst{1}Physics and Applied Mathematics Unit, Indian Statistical Institute, Kolkata 700108, India\\\inst{2}Faculty of Natural Sciences and Mathematics, University of Maribor, Koro{\v s}ka cesta 160, SI-2000 Maribor, Slovenia\\\inst{3}CAMTP -- Center for Applied Mathematics and Theoretical Physics, Mladinska 3, SI-2000 Maribor, Slovenia}


\abstract{Collective behavior among coupled dynamical units can emerge in various forms as a result of different coupling topologies as well as different types of coupling functions. Chimera states have recently received ample attention as a fascinating manifestation of collective behavior, in particular describing a symmetry breaking spatiotemporal pattern where synchronized and desynchronized states coexist in a network of coupled oscillators. In this perspective, we review the emergence of different chimera states, focusing on the effects of different coupling topologies that describe the interaction network connecting the oscillators. We cover chimera states that emerge in local, nonlocal and global coupling topologies, as well as in modular, temporal and multilayer networks. We also provide an outline of challenges and directions for future research.}

\begin{document}
\maketitle

 \section{Introduction}
Synchronizability of dynamical networks of coupled oscillators can break down into two or more synchronized and desynchronized domains when nodes in the network are connected in a nonlocal manner, and such fascinating complex spatiotemporal behavior is called a {\it chimera state} \cite{chimera_rev}.
 More precisely, the chimera state is a peculiar type of dynamical phenomenon which exhibits a hybrid structure of coexisting synchronous (coherent) and asynchronous (incoherent) behavior in a network of coupled identical oscillators with a symmetric type of coupling topology.
\par  Kuramoto and Battogtokh \cite{kuramoto} first observed the coexistence of coherence and incoherence in a network of nonlocally coupled complex Ginzburg-Landau oscillators. Later, Abrams and Strogatz \cite{strogatz} named it as chimera state and introduced some theoretical explanations for the existence of such behavior. Initially, the chimera state was investigated in phase oscillators, later  it was also found in limit-cycle oscillators \cite{hircal_chaos}, chaotic oscillators \cite{chaotic}, chaotic maps \cite{chaotic_map}, hyper chaotic time delay  systems \cite{lakshman_measure,hr_bera2} and even in neuronal systems \cite{hr_ijbc,hr_bera1} which exhibit bursting dynamics. In the beginning, chimera patterns were observed in nonlocally coupled networks and after that these states were also found  in globally (all to all coupling) \cite{global1,global2,global3}, locally (nearest neighbor) \cite{laing,hr_bera1,hr_bera2,local1} coupled networks and also in modular network \cite{chimera_modular}. Very recently, synchronous and asynchronous chimera states \cite{chimera_multiplex} have been studied between layers in the form of multiplex configuration. Chimera states have also been observed in complex networks \cite{complex_chi} and coupled oscillators with hierarchical connectivities \cite{hircal_pre,hircal_chaos}. The presence of nonlinearity in the coupling function plays a key role for the existence of chimera states in locally coupled oscillators \cite{hr_bera2,nonlinear1}. Increment in the nonlinearity of the local dynamics of nonlocally coupled Van der Pol oscillators may lead to the occurrence of multichimera states \cite{nonlinear2}. Depending on the different types of symmetry breaking situations in networks, chimera states are classified into various categories such as  amplitude-mediated chimera \cite{amc}, globally clustered chimera \cite{gcc}, amplitude chimera and chimera death \cite{cd_prl}, etc. Also based on the spatio-temporal behavior of coherent and incoherent motions, new terms are coined such as, breathing chimera \cite{breath1}, imperfect chimera \cite{imperfect_chi}, traveling chimera \cite{travelling_chi}, imperfect traveling chimera \cite{hr_bera3} and spiral wave chimera \cite{chaotic,spiral_chi}. Breathing and alternating chimera states were also observed in two coupled populations where the links between the two populations were varied with respect to time \cite{time_vry_chi} and globally clustered chimera states emerged if the links are static \cite{gcc,gcc2} over time.
\par  Noise and time delayed interactions are omnipresent in the real world systems. Robustness of chimera states have been studied in different types of networks under the impact of noise \cite{noise_chi1,noise_chi2,noise_chi3}. The research on chimera states under the influence of time delay in the coupling function of coupled systems is very important and interesting as the time delay is unavoidable due to the finite transmission speed in many physical, biological and environmental systems. Sethia et al. \cite{clustred_chi} obtained the clustered chimera states  in a nonlocally delay coupled phase oscillators and also the effect of time delay on chimera states is discussed in locally coupled networks \cite{hr_bera2} and two coupled populations \cite{gcc,gcc2}.  Beside the existence of chimera in large networks, chimera states  also emerged in a small size network \cite{small_chi} and experimentally verified in four globally coupled chaotic opto-electronic oscillators \cite{opto-elect2} and a network of four lasers \cite{laser_chi} with time delayed interaction. Apart from the manifestation of chimera state in identical oscillators, it is also observed in nonidentical oscillators. Laing et al. studied the existence and emergence of chimera states in heterogeneous networks \cite{hetero_chi1,hetero_chi2,hetero_chi3}.
\par From the above discussion on chimera states, naturally a question arises regarding the robustness and stability of it over a long time. By providing solid numerical evidence together with Lyapunov spectrum analysis, Wolfrum et al. \cite{transit_chi} revealed that chimera states are long lived chaotic transients  and later it has been proved that how chimera states persist over a long time \cite{pasre_chi}. However, Yao et al. \cite{robust_chi1} studied the robustness of chimera states against the random removal of links from the network structure. Omelchenko et al. \cite{robust_chi2} showed that the chimera states are robust with respect to the symmetric coupling topology with nonidentical oscillator and irregular coupling configuration with identical oscillators. Several theoretical and analytical approaches have been attempted for the stability analysis and characterization of the chimera states. In this context, a real-valued local order parameter \cite{chaotic_map}, Lyapunov spectrum analysis \cite{lyp_spectra} and long-time averaged mean phase velocity \cite{mean_phase} were used to characterize the chimera states. Due to the failure of distinguishing amplitude chimera from frequency chimera \cite{feqn_chi} by using the above methods, a statistical measure is developed named as strength of incoherence \cite{lakshman_measure} which characterizes various collective states from the time-series only.
\par Beside several theoretical investigations on chimera states, appearance of these states have been verified in many experimental setups, such as electronic circuits \cite{electronic}, chemical oscillator \cite{chemical_exp}, electrochemical \cite{electro}, opto-elctronic \cite{opto-elect}, mechanical \cite{mechanical} systems and  frequency modulation delay oscillators \cite{virtual_chi}.  In the real world, chimera or chimera-like behaviors are strongly connected to many man-made systems, such as power grid \cite{power1,power2}, social network \cite{social}, and also it is associated with several neuronal activity \cite{neuo}. Recently, chimera states were also observed in nonlocally coupled two-dimensional network of neuronal systems \cite{neuron_2d}.

\section{Network topology}

At the dawn of studies on chimera states, it was believed that nonlocal coupling topology is the necessary condition for the existence of chimera states. But many recent studies on chimera states explained that nonlocal coupling topology is not essential for the emergence of such states. In this perspective, we will focus on the effect of different coupling configurations for the emergence of chimera states.

\subsection{Local coupling topology}
In this section, we discuss the emergence and existence of chimera states in networks of locally coupled oscillators. At first, chimera states were detected in nonlocally coupled oscillators in the year 2002 by Kuramoto and Battogtokh \cite{kuramoto}, then after 13 years, C. Laing \cite{laing} showed the existence of chimera states in networks of purely local coupling topology. Here author considered different cases: first considering the slow-fast reaction diffusion equations in a one dimensional spatial domain where interaction was  through diffusion. Next, assuming the purely local diffusive coupling while taking non-identical complex systems into account, he found the chimera solutions where the non-uniformity was introduced by random frequencies from a Lorentzian distribution. Lastly, the author also considered a network of locally coupled identical Stuart-Landau oscillators with periodic boundary conditions in presence of nonisochronicity  parameter.  With proper numerical evidence and rigorous bifurcation analysis, author clearly articulated the existence of chimera states in locally coupled systems by considering the above three cases.
\par
\begin{figure}
	\centerline{\includegraphics[scale=0.95]{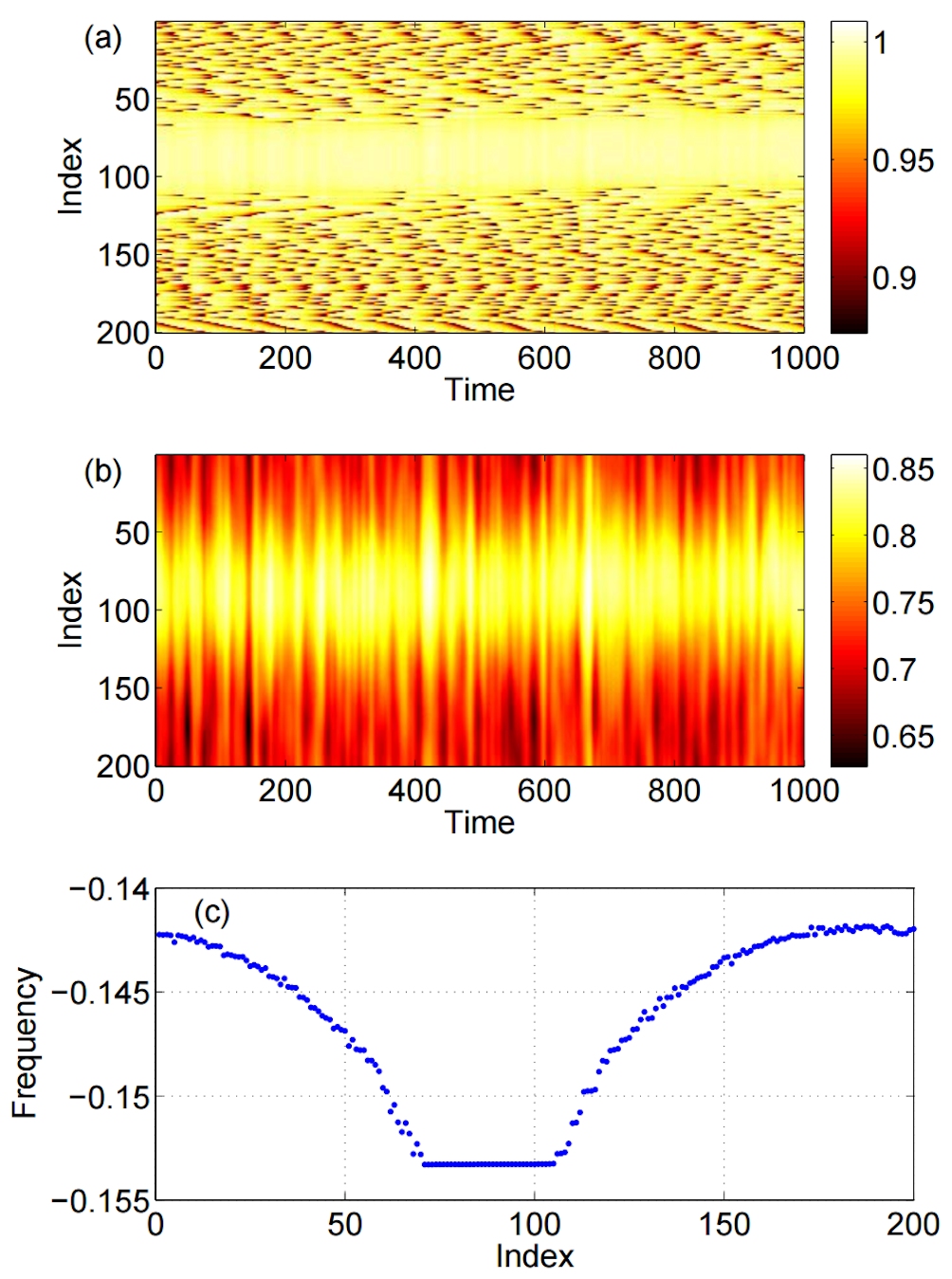}}
	\caption{The spatiotemporal dynamics of the variables (a) $|W_j|$,  and (b) $|Z_j|$ of locally coupled Landau-Stuart oscillators are plotted for fixed $C_1=-1, C_2=1, K=0.1, \epsilon=0.01, N=200$ and $\omega_0=0$. (c) The long time average of rotation frequency of $W_j$ is plotted for the confirmation of chimera sates. Figure reproduced with permission from the American Physical Society \cite{laing}.}
	\label{local}
\end{figure}

 In order to study the chimera states using local coupling, we consider $N$ Landau-Stuart oscillators interacting through other complex variables. These $N$ oscillators are equally spaced on a domain of length $1,$ with periodic boundary condition \cite{laing}. The mathematical form of the network is as follows:
\begin{equation}
\begin{array}{lcl}
\dot W_j=(1+i\omega_0)W_j-(1+iC_2)|W_j^2|W_j\;\;\;\;\;\;\;\;\;\;\;\;\;\;\;\\\;\;\;\;\;\;\;\;\;\;\;\;+K(1+iC_1)(Z_j-W_j),\\\\
\epsilon\dot Z_j=W_j-Z_j+\frac{1}{16(\Delta x)^2 } (Z_{j+1}+Z_{j-1}-2Z_j),
\end{array}
\end{equation}
where $W_j,Z_j\in \mathbb{C}$ for $j=1,...,N$ and $C_1,C_2,\epsilon,\omega_0,K$ are real parameters with $\Delta x=\frac{1}{N}.$ The long time behaviors of the variables $|W_j|$ and $|Z_j|$ are depicted in Figs. \ref{local}(a) and \ref{local}(b) respectively. Figure \ref{local}(c) shows the long time average of rotation frequency of $W_j$ which confirms the presence of a chimera state.

\par After that, chimera and multichimera states are also observed in locally coupled neuronal systems \cite{hr_bera1}, where the individual nodes of the network exhibit bursting dynamics. Here all neurons of the coupled network are considered to be identical and interacting through a chemical synaptic coupling function. The presence of the chemical synaptic coupling function in the locally coupled bursting neurons plays a  key role for the emergence of chimera states.
\par In ref. \cite{hr_bera2}, the existence of chimera states are studied in networks of locally coupled chaotic and limit cycle oscillators by taking different nonlinear coupling functions. They clearly enunciated that with suitable design of nonlinear functions in purely local coupling topology, various types of chimera states may arise. Also the effect of time delay in the coupling function is discussed. Later, Shepelev et al. \cite{nonlinear1} found the chimera state by taking an additional nonlinear unidirectional coupling function  in presence of scalar diffusive coupling.
\par The various types of chimera patterns also emerged in networks of coupled neuronal systems using local synaptic gradient coupling \cite{hr_bera3}.  In this coupling function, the synapses are excitatory or inhibitory depending on the value of two parameters, namely, gradient and synaptic coupling strengths. With proper tuning of these two coupling strengths, different types of chimera patterns are observed, such as imperfect chimeras, traveling chimeras and imperfect traveling chimeras.
 \par In this context, Hizanidis et al. \cite{local1} investigated the emergence of robust multiclustered chimera states in a dissipative driven system. In this work, each oscillators of the coupled system are taken as identical superconducting quantum interference device oscillators which are symmetrically and locally coupled. They found chimera states by properly chosen  initial conditions and observed that the extreme multistability in the coupled systems is the key feature to generate such states. Also, Clerc et al. \cite{local2} found the chimera-like states in locally coupled oscillators in presence of nonlinear damping. They also showed that the family of chimera-type states may appear  or disappear depending on initial conditions through homoclinic snaking bifurcation. Beside several theoretical and numerical studies of chimera states in locally coupled oscillators, experimentally it is verified in a chain of coupled electronic oscillators \cite{electronic}. By performing an experiment, they investigated a new state where synchronized population coexists with a spatially patterned oscillation death state.

\subsection{Nonlocal coupling topology}
Nonlocal coupling is very crucial to investigate due to its ubiquity of application in diverse fields, such as  physics, chemistry and biology etc. In this section, we deal with a brief discussion on the discoveries of different chimera patterns in nonlocally coupled oscillators.
\par
\begin{figure}
 	\centerline{\includegraphics[scale=1]{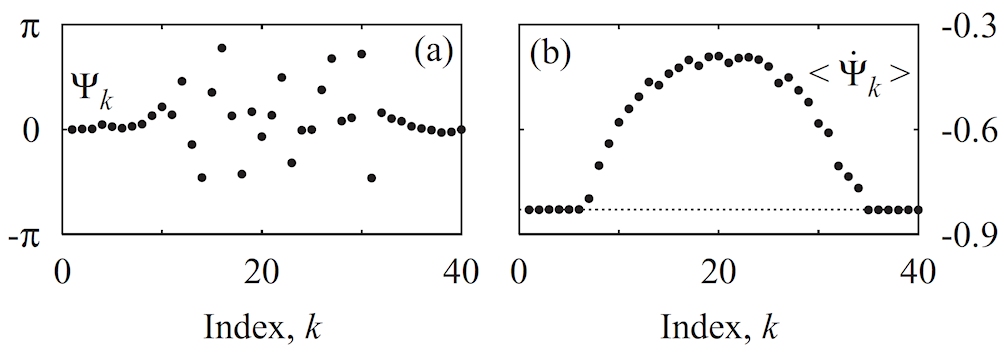}}
 	\caption{ (a) The snapshot of phase $\Psi_k$ at particular point in time, and (b) the long time averaged frequency $\langle\dot \Psi_k\rangle$  for $R=14$, $N=40$ and $\alpha=1.46$. Figure reproduced with permission from the American Physical Society \cite{transit_chi}.}
 	\label{nonlocal}
 \end{figure}

 To show the emergence of chimera states in nonlocally coupled networks, we take a ring of N identical coupled phase oscillators with nonlocal coupling topology of finite coupling range $R$ \cite{transit_chi} in the form
	\begin{equation}
	\begin{array}{lcl}
\dot \Psi_k=\omega-\frac{1}{2R}\sum_{j=k-R}^{j=k+R}\sin[\Psi_k(t)-\Psi_j(t)+\alpha],
\end{array}
\end{equation}
where $k=1, \ldots, N$ and $\omega$ is the natural frequency of the isolated oscillator and $\alpha\in(0,\frac{\pi}{2})$ is the phase-lag parameter. Figures \ref{nonlocal}(a) and \ref{nonlocal}(b) show the snapshot of phases $\Psi_k$ of Eq.~(2) and the corresponding time average frequencies $\langle\dot{\Psi}_k\rangle$, respectively. The time average frequency of the coherent group is constant and in the incoherent group they are arranged in an arc shaped profile together which signify the emergence of chimera state.

 The investigation reported in \cite{ijbcstr} revealed that the chimera pattern generates through a continuous bifurcation from a spatially modulated drift state and destroys over a saddle-node collision with an unstable version of itself. Abrams et al. \cite{breath1} produced first, the exact results regarding the stability, dynamics and bifurcations of chimera states by inspecting a model comprising of two interacting populations of oscillators expressed by
\begin{equation}
 \frac{\partial \theta_i^\sigma}{\partial t}=\omega+\sum\limits_{\sigma'=1}^{2}\frac{K_{\sigma \sigma'}}{N_{\sigma'}}\sum\limits_{j=1}^{N_{\sigma'}}sin(\theta_j^{\sigma'}-\theta_i^\sigma-\alpha),
 \end{equation}
 where $\sigma=1, 2$; $N_\sigma$ is the number of oscillators in population $\sigma$ and $K_{\sigma \sigma'}$ being the coupling strength from oscillators in $\sigma'$ to the oscillators in $\sigma$.

 \par  Apart from these studies on nonlocally connected phase oscillators, there have been a lots of analysis on chimera states in other systems having amplitude dynamics. In this context, Omelchenko et al. \cite{mean_phase} considered a ring of nonlocally coupled (with a rotational matrix form) FitzHugh-Nagumo oscillators and showed that as a result of strong coupling interaction, multichimera
 states (comprising  multiple domains of incoherence) arise. In \cite{chaotic_map}, authors conferred the disruption of spatial coherence in networks of coupled oscillators with nonlocal interaction. They diagnosed the appearance of multistable chimera-like states using coupled maps of both chaotic and periodic nature as well as with time continuous R\"{o}ssler systems. The blended feature of chimera pattern and oscillation suppression (known as "chimera death") occurring due to the co-action of nonlocality and breaking of rotational symmetry by the coupling is reported in \cite{cd_prl}. Recently, Omelchenko et al. \cite{tweezer} proposed an useful control scheme having symmetric and asymmetric control terms in order to stabilize chimera states in large and small size networks.
 
   \subsection{Global coupling topology}
This section is devoted to the study of emergence of chimera or chimera-like states in globally (all-to-all) coupled networks. Global network is the simplest and most symmetric type of network compared to nonlocal and local networks. As chimera refers to a symmetry breaking case,  so it was unexpected to find chimera patterns in globally coupled networks. Many recent results showed that chimera is not only possible using nonlocal and local coupling but it also emerges in globally coupled networks.
\par Omelchenko et al. \cite{global_prl2} studied the ensemble of globally coupled oscillators with time delayed feedback and they enunciated that chimera states appear due to spatially modulated delayed feedback. By rigorous bifurcation analysis, they affirmed that such symmetry breaking states are the natural link between the coherent and incoherent states. Later, Yeldesbay et al. \cite{global1} demonstrated the occurrence of chimera-like states in a network of identical globally coupled oscillators. They showed that presence of bistable features in the coupled systems play crucial role for the emergence of chimera states. Such bistability may occur in monostable systems with internal time delay feedback in an isolated node. Proper numerical justification using coupled Landau-Stuart systems and Kuramoto phase oscillators ensures the occurrence of chimera-like states in globally coupled networks. Recently, Chandrasekar et al. \cite{global2} examined chimera states in globally coupled oscillators  where intensity is introduced in each individual unit of the network. The mechanism behind inducing intensity is to increase the multistability of the coupled systems. They studied the effect of intensity parameter in the emergence of chimera states of globally coupled Van der Pol and chaotic R\"{o}ssler oscillators. In this case, chimera states
appeared for a well prepared initial condition due to the presence of extreme multistability.
\par Also Schmidt et al. \cite{global_prl3}  showed that a clustering mechanism is a first step for the appearance of chimera states in a network coupled via nonlinear global interaction.  Depending on the amplitude variation, they categorized various types of clusters which leads to occurrence of the different chimera states. Later Schmidt et al. \cite{global_chaos3} dealt with an oscillatory medium and obtained the chimera states with some regions displaying turbulence and the remaining portion showing the synchronized dynamics and pointed out that diffusional coupling is non-essential for such complex dynamics.

\begin{figure*}
	\centerline{\includegraphics[scale=1.12]{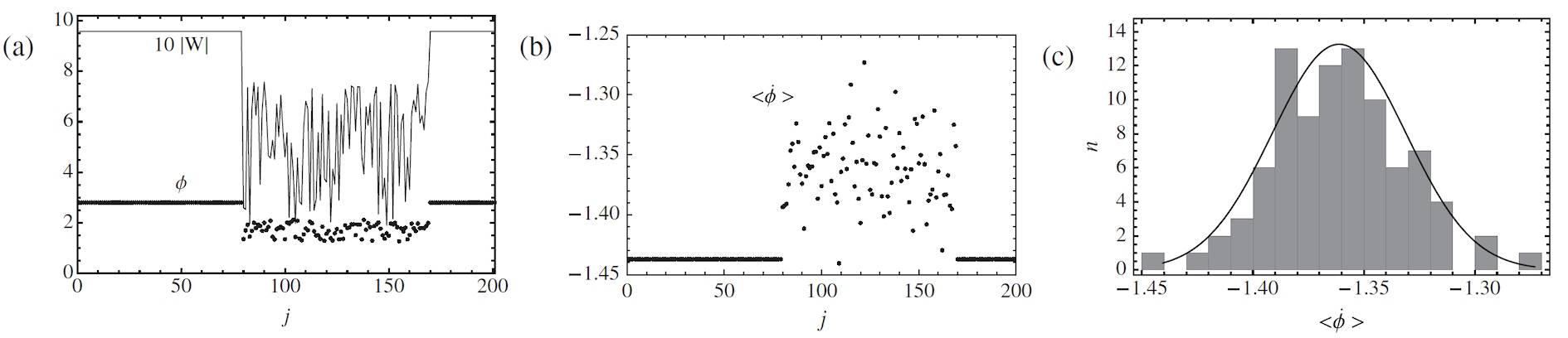}}
	\caption{Amplitude-mediated chimera state in a globally coupled network: (a) Snapshots of the profiles of the amplitude $|W|$ (multiplied by 10) and  phase $\phi$ at $K=0.70$, $C_1=-1.0$ and $C_2=2.0$. (b) Long-time average of the frequencies $\dot{\phi}$ of the oscillators. (c) A histogram of the frequencies $(\langle\dot{\phi}\rangle)$ in the incoherent segment with the corresponding Gaussian distribution. Figure reproduced with permission from the American Physical Society \cite{amc_prl}.}
	\label{bdsg}
\end{figure*}

\par Recently, Mishra et al. \cite{global3} argued that global coupling does not need to be nonlinear for the emergence of chimera states. They showed that in presence of both linear attractive and repulsive mean-field coupling functions, chimera-like states may appear in a network of globally coupled network. They identified two types of chimera-like states by taking bistable Li\'{e}nard system. The first type is chaos-chaos chimera-like states where coherent and incoherent populations of the chimera states are in chaotic motion and in other types of chimera-like states, the coherent population showed the periodic oscillation while incoherent population goes through irregular motion with small amplitude.
\par Sethia et al. \cite{amc_prl} found a new type of symmetry breaking state and named it {\it amplitude-mediated} chimera state  by considering a system of globally coupled complex Ginzburg-Landau oscillators. Here, for the first time, they observed the amplitude activity in formation of the chimera states. For this study, they considered a large population of globally coupled complex Ginzburg-Landau type identical oscillators \cite{amc_prl} whose dynamics can be modeled by the following set of equations
\begin{equation}
\dot W_j=W_j-(1+iC_2)|W_j^2|W_j+K(1+iC_1)(\bar {W_j}-W_j),
\end{equation}
where $W_j\in \mathbb{C}, \bar{W}_j=\frac{1}{N}\sum_{n=1}^{N} W_n$ for $j=1,...,N$ and $C_1, C_2, K$ are real constants.
Figure \ref{bdsg} depicts the results of amplitude-mediated chimera states in globally coupled networks (4). The snapshot of amplitudes and phases of the oscillators are shown in Fig. \ref{bdsg}(a) by solid line and dotted points, respectively. Figure \ref{bdsg}(b) displays the corresponding long time averaged frequencies $\langle \dot \phi \rangle$ of the oscillators. From this figure, the domains having coherent (constant averaged frequency) and incoherent (randomly distributed averaged frequency) dynamics in amplitude-mediated chimera state are clearly visible. Figure \ref{bdsg}(c) shows the histogram of the time average frequencies in the incoherent domain and the corresponding Gaussian distribution.

\par In ref. \cite{global4}, authors studied the emergence of chimera states in a network of globally coupled semiconductor lasers using amplitude-phase coupling with delayed optical feedback. Using random initial conditions, they found the stable chimera states in four coupled lasers and also discussed the link of multistability regime between synchronous steady-state and asynchronous periodic solutions. Wang et al. \cite{global5} studied the emergence of chimera states in frequency-weighted network of Kuramoto oscillators  with heterogeneous frequency. They found the chimera states where the oscillators having negative frequency are desynchronized and oscillators with positive natural frequency are in synchronized motion, where a weighting exponent played the key role.

\par Robustness of chimera states in small size globally coupled networks has been verified recently by an experiment. Hart et al. \cite{opto-elect2} observed the chimera and cluster states in a minimally all-to-all four coupled chaotic opto-electronic oscillators. They described that this is the minimal network size that can support chimera states and obtained some multistable region where chimera states coexist with different collective states.

\subsection{Arbitrary coupling topology}
Initially, chimera states were observed in networks of identical oscillators with  symmetric coupling topology. Recently, some studies have defeated these limitations. Chimera or  chimera-like  states were also observed in networks with  arbitrary interaction topologies. The existence of chimera-like states in modular neural networks based on the connectome of C. elegans soil worm in presence of both electrical and chemical synapses is analyzed by Hizanidis et al. \cite{chimera_modular}. Omelchenko et al. \cite {robust_chi2} affirmed that chimera states are robust against the perturbations in the form of irregular coupling topologies in networks of identical FitzHugh-Nagumo oscillators. However, they found that alterations in coupling topologies cause several qualitative changes of chimera patterns, e.g., a change of the multiplicity of incoherent domains of the chimera state. Regarding this aspect, robustness of chimera states against random removal of links in symmetrically coupled networks is reported by Yao et al. \cite {robust_chi1}.
\par
\begin{figure}
  	\centerline{
  		\includegraphics[scale=1]{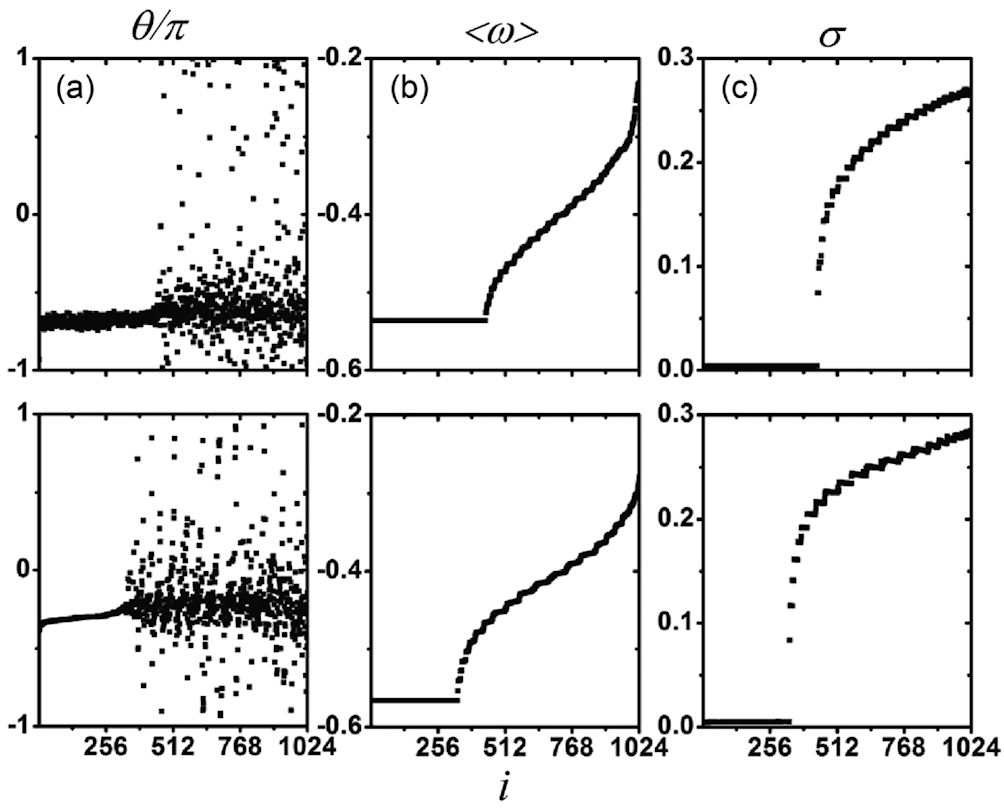}}
  	\caption{The snapshot of phases of the oscillators are plotted in column (a). Columns (b) and (c) show the corresponding effective angular velocities $\langle\omega\rangle$ of
  		oscillators  and the fluctuation of the instantaneous angular velocity $\sigma$ of oscillators respectively. The results are shown in upper and lower panel for Erd\"{o}s-R\'{e}nyi  and scale-free networks for fixed mean degree of the network $\langle k \rangle=4, N=1024, A=1, \kappa=0.1,$ $\omega=0$ and $\alpha=\frac{\pi}{2}-0.1$. Figure reproduced with permission from the American Physical Society \cite{complex_chi}.}
  	\label{random}
  \end{figure}

 In \cite{complex_chi}, authors have investigated properties of chimera states on complex networks realized by scale-free (SF) and Erd\"{o}s-R\'{e}nyi  random (ER) architectures.   We consider $N$ coupled phase oscillators which is described by the following equation
\begin{equation}
\dot \theta_i=\omega-\frac{1}{N}\sum_{j=1}^{N}G_{ij}\sin(\theta_i-\theta_j+\alpha),
\end{equation}
 for $j=1,...,N;$ $\omega$ and $\alpha$ are respectively the natural frequency and phase-lag parameter.  $G_{ij}$ is the coupling function depends on shortest length $d_{ij}$ between the $i$-th and $j$-th oscillators in the underlying complex network. The snapshots of the phases for ER (upper panel) and SF (lower panel) networks are respectively shown in Fig. \ref{random}(a).

To characterize the existence of chimera states, the effective angular velocity defined as $\langle\omega_i\rangle=\lim_{T\to\infty}\frac{1}{T} \int_{t_0}^{t_0+T} \dot\theta_i dt$ is used. Figure \ref{random}(b) show the corresponding effective angular velocity for ER (upper panel) and SF (lower panel) networks.  To further distinguish between stationary and breathing chimeras, the instantaneous angular frequency $\sigma_i^2=\lim_{T\to\infty}\frac{1}{T} \int_{t_0}^{t_0+T} (\dot\theta_i-\langle\omega_i\rangle)^2 dt$ is calculated and shown in Fig. \ref{random}(c).

\par Existence of different types of chimera states, namely stable, breathing, and alternating chimera patterns in time-varying complex networks consisting of two coupled sub-populations of Kuramoto oscillators, where inter population links are assumed to vary with time, is reported in \cite{time_vry_chi}. Chimera states caused by distance-dependent power-law coupling in networks of the realistic ecological Rosenzweig-MacArthur model are studied in \cite{tbanpower}. Transitions between spatial synchrony and different chimera patterns due to variation in the power-law exponent are discussed. Traveling multichimera states induced by hierarchical coupling topologies in nonlocally coupled lattice of limit cycle model \cite{hircal_pre} are analyzed. Moreover, another study \cite{hircal_chaos} discussed chimera patterns with various numbers of incoherent domains depending on the level of hierarchy in ring networks of Van der Pol oscillators with hierarchical coupling topology.
 \par Laing et al. \cite{hetero_chi3} reported the existence of chimera in random non-complete networks of phase oscillators and observed that these chimera states are really sensitive to elimination of connections in a random manner. Observation of metastable chimera states includes the work on this phenomenon by Shanahan et al. \cite{shancomm} in networks of community-structured phase oscillators.
 \par Recently, there has been a strong urge in detecting diverse chimera patterns in networks having multiplex (multilayer) framework. Ghosh et al. \cite{jalan1} investigated nonlocally coupled identical chaotic maps in multiplex networks and found both intra layer and layer chimera states. In this context, asynchronous and synchronous inter layer chimera states are envisaged using nonlocally coupled phase oscillators and Hindmarsh-Rose neuron models in \cite{chimera_multiplex}. Interestingly, the emergence of chimera patterns in a network of uncoupled neurons is induced by a multilayer framework having another layer of globally coupled neurons performing as the medium of interaction \cite{chimera_uncoup}.
 
\section{Conclusions and challenges ahead}
In summary, we have reviewed the existence of chimera states in complex nonlinear oscillator's networks with different types of coupling topologies, namely local, nonlocal, global and arbitrary configurations. We also discussed different types of chimera states and some recently developed methods to characterize them.  Since the discoveries on chimera states, many of the open questions have been answered during the last two decades but some important questions arise which have yet to be answered clearly. Furthermore, new challenges for chimera states appear in different fields. These states may be extended in real-world networks, e.g. power grid, human brain, food webs etc. where such behavior has not been understood clearly. Fast discoveries on chimera states lead some challenges issues. For example, first challenging issue is how to control the coherent and incoherent subpopulations in the chimera states. Although, some research works have been done regarding this controlling issue but still there is no universal and suitable technique invented. Bick et al. \cite{control_njp} prescribed a strategy to control the chimera state where their scheme is based on gradient dynamics to exploit drift of a chimera. Using this controlling technique they dynamically modulated the desired position of coherent population in chimera states. Semenov et al. \cite{control_aipconf} studied the controlling issue of chimera states using deterministic and stochastic external periodic forcing. Recently, Gambuzza et al. \cite{control_pinning} presented a control scheme of chimera states based on pinning control  in a system of nonlocally coupled FitzHugh-Nagumo and Kuramoto oscillators. They articulated that by proper pinning control, coherent and incoherent population of chimera states can be suitably monitored. Very recently, a controlling technique of chimera states in nonlocally coupled ring-networks of FitzHugh-Nagumo elements is mentioned \cite{control_excite}. They studied the influence of excitability of few nodes in the network, and even they showed that one excitable element with all the other nodes in oscillatory motion in the network is sufficient to control the coherent and incoherent dynamics in  chimera states. Most of the above discussed control protocol of chimera states is studied in nonlocal coupling topology but there is no control strategy, to the best knowledge of us, in local, global and other type of coupling configuration in the literature. So, now a days, the development of general mechanism for controlling chimera states irrespectively of coupling topology is an important issue.
\par Next, the role of initial conditions of each oscillators in the networks for the emergence of chimera states is another challenging issue. From the discoveries, it was a fundamental problem was how to chose initial condition for the emergence of chimera states. In this context, Martens et al. \cite{ba_chi} tracked the basin of attraction for chimera states  but they did not give any kind of quantification measure for checking the robustness of various dynamical states (coherent, incoherent, chimeras and clustered states etc.) against their initial condition. Recently, Rakshit et al. \cite{bs_chimera} investigated using basin stability analysis that there exist finite windows in the coupling strength for which only a fraction of randomly picked initial conditions lead to a chimera state. So, further precise studies on effect of initial conditions are needed for better understanding of the underlying mechanism of chimera states in different networks.
\par Also, the effect of the phase-lag and of the nonisochronicity parameter for the appearance of chimera states is an interesting issue.  Chimera states were first found in nonlocally coupled phase oscillators in presence of the phase-lag parameter, after that many studies have been done but still it is not clear how the phase-lag parameter affects the dynamics of chimera state from an analytical aspect.  Also, investigation of the role of the nonisochronicity parameter, a characteristic feature of many paradigmatic limit cycle systems, on the emergence of chimera states in globally coupled oscillators is an important issue.
\par In the case of stability and characterization of chimera states, few methods were developed such as the Ott-Antonsen approach, strength of incoherence, mean phase velocity, and local order parameter, but there is no concrete mathematical treatment which is still an open question in the chimera literature. As the research on chimera states is going so fast over time, it is not easy to present a complete scenario regarding the chimera states in dynamical networks.

\begin{acknowledgments}
D.G. was supported by the Department of Science and Technology of the Government of India (Grant EMR/2016/001039). M.P. was supported by the Slovenian Research Agency (Grants J1-7009 and P5-0027).
\end{acknowledgments}


\begin{thebibliography}{abc}
		\bibitem{chimera_rev} Panaggio M. J. \textit{et al., Nonlinearity,} {\bf 28} (2015) R67.
		\bibitem{kuramoto} Kuramoto Y. \textit{et al., Nonlinear Phenom. Complex Syst.,} {\bf 5} (2002) 380.
		\bibitem{strogatz} Abrams D. M. \textit{et al., Phys. Rev. Lett.,} {\bf 93} (2004) 174102.
		\bibitem{hircal_chaos} Ulonska S. \textit{et al., Chaos,} {\bf 26} (2016) 094825.
		\bibitem{chaotic} Gu C. \textit{et al., Phys. Rev. Lett.,} {\bf 111} (2013) 134101.
		\bibitem{chaotic_map} Omelchenko I. \textit{et al., Phys. Rev. Lett.,} {\bf 106} (2011) 234102.
		\bibitem{lakshman_measure} Gopal R. \textit{et al., Phys. Rev. E,} {\bf 89} (2014) 052914.
		\bibitem{hr_bera2} Bera B. K. \textit{et al., Phys. Rev. E,} {\bf 93}  (2016) 052223.
		\bibitem{hr_ijbc} Hizanidis J. \textit{et al., Int. J. Bifurcat. Chaos,} {\bf 24} (2014) 1450030.
		\bibitem{hr_bera1} Bera B. K. \textit{et al., Phys. Rev. E,} {\bf 93} (2016) 012205.
		\bibitem{global1} Yeldesbay A. \textit{et al., Phys. Rev. Lett.,} {\bf 112}  (2014) 144103.
		\bibitem{global2} Chandrasekar V. K. \textit{et al., Phys. Rev. E,} {\bf 90}  (2014) 062913.
		\bibitem{global3} Mishra A. \textit{et al., Phys. Rev. E,} {\bf 92}  (2015) 062920.
		\bibitem{laing} Laing C. R., \textit{Phys. Rev. E,} {\bf 92}  (2015) 050904(R).
		\bibitem{local1} Hizanidis J. \textit{et al., Phys. Rev. E,} {\bf 94}  (2016) 032219.
		\bibitem{chimera_modular} Hizanidis J. \textit{et al., Sci. Rep.,} {\bf 6}  (2016) 19845.
		\bibitem{chimera_multiplex} Maksimenko V. A. \textit{et al., Phys. Rev. E,} {\bf 94}  (2016) 052205.
		\bibitem{complex_chi} Zhu Y. \textit{et al., Phys. Rev. E,} {\bf 89} (2014) 022914.
		\bibitem{hircal_pre} Hizanidis J. \textit{et al., Phys. Rev. E,} {\bf 92} (2015) 012915.
		\bibitem{nonlinear1} Shepelev I. A. \textit{et al., Commun. Nonlinear Sci. Numer. Simulat.,} {\bf 44} (2017) 277.
		\bibitem{nonlinear2} Omelchenko I. \textit{et al., Chaos,} {\bf 25} (2015) 083104.
		\bibitem{amc} Sethia G. C. \textit{et al., Phys. Rev. E,} {\bf 88}  (2013) 042917.
		\bibitem{gcc} Sheeba J. H. \textit{et al., Phys. Rev. E,} {\bf 79}  (2009) 055203(R).
		\bibitem{cd_prl} Zakharova A. \textit{et al., Phys. Rev. Lett.,} {\bf 112}  (2014) 154101.
		\bibitem{breath1} Abrams D. M. \textit{et al., Phys. Rev. Lett.,} {\bf 101}  (2008) 084103.
		\bibitem{imperfect_chi} Kapitaniak T. \textit{et al., Sci. Rep.,} {\bf 4}  (2014) 6379.
		\bibitem{travelling_chi} Xie J. \textit{et al., Phys. Rev. E,} {\bf 90}  (2014) 022919.
		\bibitem{hr_bera3} Bera B. K. \textit{et al., Phys. Rev. E,} {\bf 94}  (2016) 012215.
		\bibitem{spiral_chi} Li B. W. \textit{et al., Phys. Rev. E,} {\bf 93}  (2016) 020202(R).
		\bibitem{time_vry_chi} Buscarino A. \textit{et al., Phys. Rev. E,} {\bf 91} (2015) 022817.
		\bibitem{gcc2} Sheeba J. H. \textit{et al., Phys. Rev. E,} {\bf 81} (2010) 046203.
		\bibitem{noise_chi1} Loos S. A. M. \textit{et al., Phys. Rev. E,} {\bf 93} (2016) 012209.
		\bibitem{noise_chi2} Zakharova A. \textit{et al., arXiv} 1611.03432 (2016).
		\bibitem{noise_chi3} Laing C. R., \textit{Chaos,} {\bf 22} (2012) 043104.
		\bibitem{clustred_chi} Sethia G. C. \textit{et al., Phys. Rev. Lett.,} {\bf 100} (2008) 144102.
		\bibitem{small_chi} Maistrenko Y. \textit{et al.,Phys. Rev. E,} {\bf 95} (2017) 010203(R).
		\bibitem{opto-elect2} Hart J. D. \textit{et al., Chaos,} {\bf 26} (2016) 094801.
		\bibitem{laser_chi} R\"{o}hm A. \textit{et al., Phys. Rev. E,} {\bf 94} (2016) 042204.
		\bibitem{hetero_chi1} Laing C. R., \textit{Chaos,} {\bf 19} (2009) 013113.
		\bibitem{hetero_chi2} Laing C. R., \textit{Physica D,} {\bf 238} (2009) 1569.
		\bibitem{hetero_chi3} Laing C. R. \textit{et al., Chaos,} {\bf 22} (2012) 013132.
		\bibitem{transit_chi} Wolfrum M. \textit{et al., Phys. Rev. E,} {\bf 84} (2011) 015201(R).
		\bibitem{pasre_chi} Suda Y. \textit{et al., Phys. Rev. E,} {\bf 92} (2015) 060901(R).
		\bibitem{robust_chi1} Yao N. \textit{et al., Sci. Rep.,} {\bf 3} (2013) 3522.
		\bibitem{robust_chi2} Omelchenko I. \textit{et al., Phys. Rev. E,} {\bf 91} (2015) 022917.
		\bibitem{lyp_spectra} Wolfrum M. \textit{et al., Chaos,} {\bf 21} (2011) 013112.
		\bibitem{mean_phase} Omelchenko I. \textit{et al., Phys. Rev. Lett.,} {\bf 110} (2013) 224101.
		\bibitem{feqn_chi} Gopal R. \textit{et al., Phys. Rev. E,} {\bf 91} (2015) 062916.	
		\bibitem{electronic} Gambuzza L. V. \textit{et al., Phys. Rev. E,} {\bf 90} (2014) 032905.
		\bibitem{chemical_exp} Tinsley M. R. \textit{et al., Nat. Phys.,} {\bf 8} (2012) 662.
		\bibitem{electro} Wickramasinghe M. \textit{et al., PLoS ONE,} {\bf 8} (2013) e80586.
		\bibitem{opto-elect} Hagerstrom A. \textit{et al., Nat. Phys.,} {\bf 8} (2012) 658.
		\bibitem{mechanical} Martens E. A. \textit{et al., Proc. Natl. Acad. Sci. USA,} {\bf 110} (2013) 10563.
		\bibitem{virtual_chi} Larger L. \textit{et al., Phys. Rev. Lett.,} {\bf 111} (2013) 054103.
		\bibitem{power1} Motter A. E. \textit{et al., Nat. Phys.,} {\bf 9} (2013) 191.
		\bibitem{power2} D\"{o}rfler F. \textit{et al., Proc. Natl. Acad. Sci. USA,} {\bf 110(6)} (2013) 2005.
		\bibitem{social} Gonz\'{a}lez-Avella J. C. \textit{et al., Physica A,} {\bf 399} (2014) 24.
		\bibitem{neuo} Levy R. \textit{et al., J. Neurosci.,} {\bf 20} (2000) 7766.
		\bibitem{neuron_2d} Schmidt A. \textit{et al., Phys. Rev. E,} {\bf 95} (2017) 032224.
				

		\bibitem{local2} Clerc M. G. \textit{et al., Phys. Rev. E,} {\bf 93} (2016) 052204.
		\bibitem{ijbcstr} Abrams D. M. \textit{et al., Int. J. Bifurcat. Chaos,} {\bf 16} (2006) 21.
		
		\bibitem{tweezer} Omelchenko I. \textit{et al., Phys. Rev. Lett.,} {\bf 116} (2016) 114101.
		\bibitem{global_prl2} Omelchenko O. E. \textit{et al., Phys. Rev. Lett.,} {\bf 100} (2008) 044105.
		\bibitem{global_prl3} Schmidt L. \textit{et al., Phys. Rev. Lett.,} {\bf 114} (2015) 034101.
		\bibitem{global_chaos3} Schmidt L. \textit{et al., Chaos,} {\bf 25} (2015) 064401.
		\bibitem{amc_prl} Sethia G. C. \textit{et al., Phys. Rev. Lett.,} {\bf 112} (2014) 144101.
		\bibitem{global4} B\"{o}hm F. \textit{et al., Phys. Rev. E,} {\bf 91} (2015) 040901(R).	
		\bibitem{global5} Wang H. \textit{et al., Phys. Rev. E,} {\bf 83} (2011) 066214.

		\bibitem{tbanpower} Banerjee T. \textit{et al., Phys. Rev. E,} {\bf 94} (2016) 032206.
		\bibitem{shancomm} Shanahan M., \textit{Chaos,} {\bf 20} (2010) 013108.
		\bibitem{jalan1} Ghosh S. \textit{et al., EPL,} {\bf 115} (2016) 60005.
		\bibitem{chimera_uncoup} Majhi S. \textit{et al., Sci. Rep.,} {\bf 6}  (2016) 39033.	
		\bibitem{control_njp} Bick C. \textit{et al., New J. Phys.,} {\bf 17} (2015) 033030.
		\bibitem{control_aipconf} Semenov V. \textit{et al., AIP Conf. Proc.,} {\bf 1738} (2016) 210013.
		\bibitem{control_pinning} Gambuzza L. V. \textit{et al., Phys. Rev. E,} {\bf 94} (2016) 022306.
		\bibitem{control_excite} Isele T. \textit{et al., Phys. Rev. E,} {\bf 93} (2016) 022217.
		\bibitem{ba_chi} Martens E. A. \textit{et al., New J. Phys.,} {\bf 18} (2016) 022002.	
		\bibitem{bs_chimera} Rakshit S. \textit{et al., arXiv} 1704.05301 (2017).
	
	
\end{thebibliography}
\end{document}